\documentclass[aps,pra,reprint]{revtex4-1}
\usepackage{times}
\usepackage{epsfig}

\begin{document}
\title{Access via laboratory experiments to the level mixing effect induced by blackbody radiation and its influence on the cosmological hydrogen recombination problem}

\author{T. Zalialiutdinov$^{1}$, D. Solovyev$^{1}$, L. Labzowsky$^{1,2}$ and G. Plunien$^{3}$}
\affiliation{ 
$^{1}$ Department of Physics, St. Petersburg State University, Petrodvorets, Ulianovskaya 1, 198504, St. Petersburg, Russia \\
$^{2}$ Petersburg Nuclear Physics Institute, 188300, Gatchina, St. Petersburg, Russia \\
$^{3}$ Institut f\"{u}r Theoretische Physik, Technische Universit\"{a}t Dresden, Mommsenstra\ss e 13, D-10162, Dresden, Germany}

\begin{abstract}
Two different effects of the blackbody radiation (BBR)-induced atomic line broadening are compared. The first one (stimulated Raman scattering) was discussed by many authors, the second one (quadratic level mixing) was predicted earlier in our publication. It is shown that the mixing effect gives the most significant contribution to the line broadening and it is indicated how to distinguish these two effects in laboratory experiments. The influence of the level mixing on the recombination history of primordial plasma is also discussed. 
\end{abstract}
\maketitle

The influence of external fields on atomic characteristics is still one of the interesting subjects for investigations in modern atomic physics. In particular a question about the blackbody radiation (BBR) influence on atoms is widely discussed. First the BBR induced effects were observed experimentally and then the theoretical description was given in \cite{gallagher,farley} within the frameworks of quantum mechanical (QM) approach. In particular, it was shown that the blackbody radiation induces the ac-Stark shift of energy levels and an additional line broadening in atoms. Theoretical calculations of the dynamic Stark shifts and depopulation rates of Rydberg energy levels caused by the BBR and the corresponding experimental measurements were widely discussed in literature \cite{porsev}-\cite{middelman}. The most important consequence of these investigations corresponds to the improvement of atomic clocks and the development of optical standards of frequency measurements \cite{lisdat}.

Finally, in \cite{solovyev} the effect of level mixing induced by the blackbody radiation was firstly described theoretically within the rigorous quantum electrodynamic (QED) theory. The mixing effect for the states of opposite parity arising in the presence of an external electric field leads to a significant changes of the decay rates, see, for example, \cite{sslg,solovyeva}. We should note that all  effects in the presence of the BBR are similar to the phenomena which take place in an external electric field. Similar to the Stark (static or dynamic) effect in the presence of 'ordinary' external electric field the energy shift of atomic levels induced by the BBR can be estimated with the use of root-mean square value of the field strength of thermal radiation (in a.u.): 
\begin{eqnarray}
\label{1}
\langle E^2(\omega) \rangle = \frac{8\alpha^3}{\pi}\omega^3 n_\beta(\omega) = \frac{8\alpha^3}{\pi}\frac{\omega^3}{e^{\beta\omega}-1},
\end{eqnarray}
where $\langle E^2(\omega) \rangle$ is rms electric field strength, $\omega$ is the radiation frequency. The Planck's distribution function is presented by $n_\beta(\omega)$ with $\beta = 1/k_B T$, $k_B$ is the Boltzman's constant, $T$ is the temperature in Kelvin and $\alpha$ is the fine structure constant. Then the integral rms value of electric field is
\begin{eqnarray}
\label{2}
\langle E^2 \rangle = \frac{1}{2}\int\limits_0^\infty \langle E^2(\omega) \rangle d\omega = \frac{4\pi^3}{15}\alpha^3 (k_B T)^4 
\\
\nonumber
= (8.319\;\mathrm{V/cm})^2\left[T(\mathrm{K})/300\right]^4.
\end{eqnarray}

In conjunction with the expression (\ref{2}) the level mixing effect induced by the thermal radiation can be introduced. The level mixing effect in an external electric field was considered in connection with the Lamb shift measurements in hydrogen and hydrogen-like ions \cite{drakebull,marrus} and the corresponding theoretical analysis of the electric field influence on atomic levels can be found in \cite{mohr,hilley}. An accurate description of level mixing effect in hydrogen atom was given in \cite{Ans}. In particular, the authors of \cite{Ans} have shown that the mixing of $2s$ and $2p$ states in hydrogen atom can mimic the parity non-conservation phenomenon. 

As a result of level mixing effect in presence of an external electric field \cite{Ans,mohr} or in presence of the BBR \cite{solovyev} the essential modification of the $2s$ state decay in hydrogen atom arises. This is due to the appearance of the one-photon electric dipole decay channel which was forbidden by the selection rules in absence of an external field. As a consequence the $2s$ state level in hydrogen atom does not remain a metastable one in presence of an external field. 

We should note that in what concerns level mixing we include only $ 2p_{1/2} $ and $ 2s_{1/2} $ mixing and neglect $ 2p_{3/2} $ and $ 2s_{1/2} $ mixing. The reason is that the fine structure interval $ E_{2p_{3/2}}-E_{2s_{1/2}} $ is much larger than the Lamb shift $ E_{2s_{1/2}}-E_{2p_{1/2}} $ and consequently the mixing effect for $ 2p_{3/2} $ state is much smaller.

The one-photon decay rate of the mixed $\overline{2s}$ state \cite{Ans,sslg} can be expressed as
\begin{eqnarray}
\label{3}
W_{\overline{2s} 1s}^{(1\gamma)}(\textbf{k}) = W_{2s 1s}^{(1\gamma)}(\textbf{k})\left[1+e a_0{\bf E}\,{\bf n_k} \frac{|\eta|^2\Gamma_{2p}}{\mathit{w}}\right. \\\nonumber \left. + e^2 a^2_0 \frac{|\eta|^2{\bf E}^2}{\mathit{w}^2}\right],
\end{eqnarray}
where ${\bf E}$ represents the electric field, ${\bf n_k}$ is the unit vector corresponding to the wave vector $\textbf{k}$ of photon, $\mathit{w} = \sqrt{W_{2s\,1s}^{\mathrm{M1}}/W_{2p\, 1s}^{\mathrm{E1}} }$, the electron charge $e$ and the Bohr's radius $a_0$ are written explicitly for clarity. $\Gamma_{2p}$ is the $2p$ level width and $\eta = \left(\Delta E^L_{2p2s}-\frac{\mathrm{i}}{2}\Gamma_{2p}\right)^{-1}$. $\Delta E^L_{2p2s}$ represents the Lamb shift between $2s_{1/2}$ and $2p_{1/2}$ levels, the one-photon transition probabilities $W_{2s\,1s}^{\mathrm{M1}}$ and $W_{2p\,1s}^{\mathrm{E1}}$ correspond to the emission of the magnetic dipole and electric dipole photons, respectively. Integration over photon emission direction ${\bf n_k}$ and frequency of the emitted photon $\omega=|\textbf{k}|$ yields the expression  \cite{solovyeva,sslg}
\begin{eqnarray}
\label{4}
W_{\overline{2s}\, 1s}^{(1\gamma)} = W_{2s\, 1s}^{\mathrm{M1}} + \frac{e^2a_0^2\mathrm{E}_{0}^2}{ (\Delta E^L_{2p2s})^2+\frac{1}{4}\Gamma_{2p}^2}W_{2p1s}^{\mathrm{E1}},
\end{eqnarray}
where $ \mathrm{E}_{0} $ is the field amplitude. This expression shows that the additional one-photon electric dipole emission channel is allowed for the hydrogen-like atom in the metastable $2s$ state in presence of an external electric field. The term linear in the field in Eq. (\ref{3}) vanishes after the integration over photon emission directions. In contrast, the term quadratic in the field does not depend on the photon emission or field directions. This contribution represents the quadratic mixing effect.
 
Since the decay rate of the $\mathrm{E1}$ transition, $W_{2p1s}^{\mathrm{E1}} = 6.265\times 10^8$ $\mathrm{s}^{-1}$, exceeds strongly the the one-photon magnetic decay channel,  $W_{2s1s}^{\mathrm{M1}} = 2.496\times 10^{-6}$ $\mathrm{s}^{-1}$, the second term in Eq. (\ref{4}) may become the dominant decay channel of the mixed $\overline{2s}$ state with increasing strength of the external electric field. The contribution of second term in (\ref{4}) at the field strength $475$ V/cm (easily achievable in laboratory experiments) becomes equal to the decay rate of the $2p$ level in hydrogen atom (the case of complete mixing) and is much larger than the main two-photon $\mathrm{E1E1}$ decay rate of the $2s$ state in absence of the electric field, $W_{2s1s}^{\mathrm{E1E1}} = 8.229$ $\mathrm{s}^{-1}$. In turn, the same scenario can be considered for the mixing of $\overline{2p}$ state in hydrogen atom. In this case there is no essential difference in the decay rate in external field because of the small additional contribution of the transition rates of the $2s$ level. Note, that this effect arises in presence of static electric field. According to the description above the rms value $ \langle E^2\rangle $ of electric field caused by the BBR can be estimated by Eq. (\ref{2}). Thus, the effect of level mixing should arise in presence of thermal radiation as it was demonstrated in \cite{solovyev}.

However, the thermal radiation can not be described completely as a static electric field. The significant dynamical character of BBR modifies the form of the transition rate.

The full description of the dynamical effects as well as mixing effects was given in \cite{solovyev}. The QED expression for the BBR broadening is

\begin{eqnarray}
\label{5}
\Gamma_a^{\mathrm{BBR-QED}} = \frac{2e^2}{3\pi}\sum\limits_{n}
|\langle a |{\bf r}| n \rangle|^2
\int\limits_{0}^{\infty}d\omega n_\beta(\omega) \omega^3
\times
\\
\nonumber
\left[\frac{\Gamma_{na}}{(\tilde{\omega}_{na}+\omega)^2 + \frac{1}{4}\Gamma_{na}^2} + \frac{\Gamma_{na}}{(\tilde{\omega}_{na}-\omega)^2 + \frac{1}{4}\Gamma_{na}^2}\right],
\end{eqnarray}
where $a$ and $n$ denote the set of quantum numbers of corresponding atomic state, $\tilde{\omega}_{na}\equiv E_{n}-E_{a}+\Delta E^L_{na}$, $\Delta E^L_{na}$ is the corresponding Lamb shift and $\Gamma_{na}\equiv \Gamma_n+\Gamma_a$.  Expression (\ref{5}) is the width of resonant emission line profile in presence of BBR. 

For the dynamical effects in Eq. (\ref{5}) the frequency-dependent energy denominators are responsible. The mixing effect is incorporated when the summation over $ n $ in Eq. (\ref{5}) extends over the states with the space parity opposite to the parity of the state $ a $. In this case the main contribution comes from the state $ n $ close to $ a $. In hydrogen atom such states ($ 2s_{1/2} $ and $ 2p_{1/2} $ for example) are degenerate and $ \omega_{an}=\Delta E^{L}_{an} $ is the Lamb  shift. For such levels as $ 2s $ in hydrogen which will be of our interest below, the mixing effect becomes dominant. 

To clarify the physical situation we have to compare the result Eq. (\ref{5}) with the well-known effect of the level broadening by the multiple photon scattering (Raman scattering in general case) on atomic levels: $i + \gamma \rightarrow a \rightarrow f + \gamma'$, where $i$, $f$ denote the initial and final states, respectively, $a$ is the excited intermediate state and $\gamma$ represents the emitted/absorbed photon. In case of BBR this effect was described in \cite{farley} and has application in cosmological recombination \cite{seager2000}. Below we will show that 1) the mixing broadening can not be reduced to the Raman scattering (RS); it is independent effect 2) mixing effect dominates over RS in line broadening 3) mixing effect leads to the emission lines that occur at frequencies Lamb shifted from frequencies corresponding to the emission lines broadened by the RS process 4) This frequency difference in principle is possible to observe in laboratory experiments what may give an access via these experiments to the study of primordial plasma.

For this purpose below we consider the simple RS process. The multiple photon scattering will be taken into account by introducing the Einstein coefficients, or the number of photons. In case of the BBR the number of photons is defined by the Planck's distribution function. 

The S-matrix element of the RS process can be written in the form \cite{dimaNR}-\cite{theory}:
\begin{eqnarray}
\label{7}
\hat{S}^{(2)}_{fi}=(-\mathrm{i}e)^2\int dx_1dx_2\overline{\psi}_{f}(x_1)\gamma_{\mu_1}A^{*(\textbf{k}_2,\textbf{e}_2)}_{\mu_1}(x_1)
\\
\nonumber
\times S(x_1x_2)\gamma_{\mu_2}A^{(\textbf{k}_1,\textbf{e}_1)}_{\mu_2}(x_2)\psi_{i}(x_2),
\end{eqnarray}
where $\psi_{i}(x)$ and Dirac conjugated $\overline{\psi}_{f}(x)$ represent the wave functions of the initial and final states, respectively, $\gamma_{\mu}$ are the Dirac matrices with $\mu={0,1,2,3}$. The photon wave function (electromagnetic field potential) is described by
\begin{eqnarray}
\label{8}
A_{\mu}^{(\textbf{k},\textbf{e})}(x)=\sqrt{\frac{2\pi}{\omega}}e^{\;(\lambda)}_{\mu}e^{\mathrm{i}k_{\mu}x_{\mu}}=A_{\mu}^{(\textbf{k},\textbf{e})}(\textbf{r})\;e^{-\mathrm{i}\omega t},
\end{eqnarray}
where $k\equiv(\textbf{k},\omega)$ is the photon momentum 4-vector, $\textbf{k}$ is the photon wave vector, $\omega = |\textbf{k}|$ is the photon frequency, $e^{\;(\lambda)}_{\mu}$ are the components of the photon polarization 4-vector. $A_{\mu}^{(\textbf{k},\textbf{e})}$ and $A_{\mu}^{*\;(\textbf{k},\textbf{e})}$ in (\ref{7}) correspond to the absorbed and emitted photon, respectively. In the Furry picture the eigenmode decomposition of Feynman electron propagator $S(x_1, x_2)$ reads \cite{akhiezer}
\begin{eqnarray}
\label{9}
S(x_1x_2)=\frac{1}{2\pi \mathrm{i}}\int\limits_{-\infty}^{\infty}d\omega\;e^{\mathrm{i}\Omega_1(t_1-t_2)}\sum\limits_n\frac{\psi_n(\textbf{r}_1)\bar{\psi}_n(\textbf{r}_2)}{E_n(1-\mathrm{i}0)+\omega},\,\,\,\,
\end{eqnarray}
where the summation over $ n $ in Eq. (\ref{9}) extends over the entire Dirac spectrum.

The differential absolute probability of emission process resulting from RS cross-section in case of resonant scattering is \cite{akhiezer, lab1994}
\begin{eqnarray}
\label{12}
d{\it w}_{af}(\omega)=\frac{1}{2\pi}\frac{dW_{af}(\omega)}{(E_a-E_f-\omega)^2+\frac{1}{4}\Gamma_{a}^2},
\end{eqnarray}
where $dW_{af}$ is the differential partial transition rate $a\rightarrow f$.

Expression (\ref{12}) follows from separation of absorption and emission processes, which become independent within the resonant approximation. The result (\ref{12}) represents the emission line profile, i.e. the photon emission occurs at the resonant frequency $\omega_{af}=E_a-E_f $. Integrating Eq. (\ref{12}) over frequency $ \omega $ we find
\begin{eqnarray}
\label{15}
{\it w}_{af}=\frac{W^{\mathrm{E1}}_{a f}(\omega_{a f})}{\Gamma_a},
\end{eqnarray}
where ${\it w}_{af}$ is the absolute transition probability $a\rightarrow f$ and $W^{\mathrm{E1}}_{a f}(\omega_{a f})$ denotes the electric dipole one-photon spontaneous emission rate. Expression (\ref{15}) is given for the case of one-photon emission process that arises as a result of Raman scattering in the resonanace approximation. However, the presence of a photon field (in particular BBR) induces additional emission with the same frequency \cite{landauqed}. According to \cite{landauqed} an induced photon emission probability is expressed via the number of photons. In our case this number is defined by the Planck's distribution function, $n_{\beta}$:
\begin{eqnarray}
\label{16}
W^{\rm ind}_{a f}= n_{\beta}\left(\omega_{a f}\right) W^{\mathrm{E1}}_{a f}(\omega_{a f}).
\end{eqnarray}
Thus, the emission probability corresponding to the stimulated RS process is given by
\begin{eqnarray}
\label{18}
W_{a f}(T) =  \left(1+n_{\beta}\left(\omega_{a f}\right)\right)W^{\mathrm{E1}}_{a f}(\omega_{a f}).
\end{eqnarray}
Then the total BBR-induced level broadening via RS process for an arbitrary level $ a $ in the nonrelativistic limit (neglecting all the types of photons except $ \mathrm{E1} $) looks like
\begin{eqnarray}
\label{19}
\Gamma^{\rm BBR-RS}_a
= \frac{4e^2}{3}\sum\limits_{f} \left|\langle a | \textbf{r} |f\rangle\right|^2\frac{\omega_{af}^3}{e^{\beta\omega_{af}}-1},
\end{eqnarray}
and coincides precisely with the result obtained in \cite{farley} within the QM approach. Here we have used an explicit expression for $W^{\mathrm{E1}}_{a f}$. Expression (\ref{19}) is the sum of all the partial transition probabilities including the higher excited states. The one-photon emission occurs at the corresponding resonant frequency $\omega_{af}$.

Thus, we have two effects induced by the BBR: Raman scattering and level mixing. They have similar structure. The former one is described by the expression (\ref{19}).  The expression (\ref{5}) is more general, it incorporates both effects. The result (\ref{19}) can be obtained from Eq. (\ref{5}) if we fully neglect the widths of the states $ n,a $. The limit $\Gamma_{na}\rightarrow 0$ in square brackets in Eq. (\ref{5}) gives the sum of two delta functions  $\delta\left(\tilde{\omega}_{na}+\omega\right)+\delta\left(\tilde{\omega}_{na}-\omega\right)$ and, therefore, the RS result (\ref{19}) arises immediately. 

The QM mixing effect follow from Eq. (\ref{4}) by substituting the rms value Eq. (\ref{2}) instead of $ \mathrm{E}_{0}^{2} $. This result, also can be obtained from Eq. (\ref{5}) but using different approximations than for RS effect. To do this we have to set $ \omega=0 $ in square brackets in Eq. (\ref{5}). It can be justified if we remember that function $ n_{\beta}(\omega) $ is concentrated at small $ \omega $ values for not too high temperatures. Moreover Eq. (\ref{4}) follows from Eq. (\ref{5}) when we take the contribution $\Gamma_n $ from the sum $ \Gamma_{na}=\Gamma_n+\Gamma_a $ in the numerators in square brackets in Eq. (\ref{5}). The contribution of $ \Gamma_{a} $ also appears to be important, but  it corresponds not to the mixing but to the general 'dynamic' effect.   

\begin{figure}
\centering
\caption{\small{Emission line profiles Eq. (\ref{20}) for the transitions $\overline{2p}\rightarrow 1s +\gamma(\mathrm{E1})$ (yellow-dashed line) and $\overline{2s}\rightarrow 1s +\gamma(\mathrm{E1})$ (blue-solid line) at the BBR temperature $T=3000$ K. The magnitudes of emission line profiles and frequency interval are normalized to unity. The Ly$_\alpha$ frequency corresponds to $\omega=0.5$. The corresponding values of transition rates and level widths are presented in Table~\ref{width1}.}  }
\includegraphics[scale=0.5]{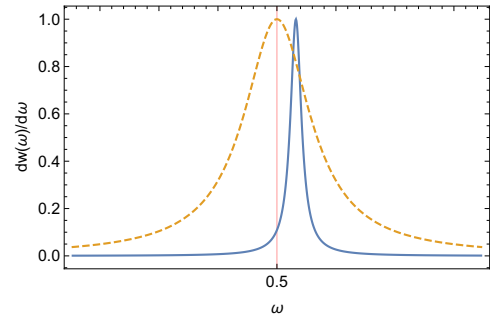}
\label{fig1}
\end{figure}

According to \cite{mohr,Ans} the mixing of atomic levels with opposite parity occurs in a static electric field. Such a field can not induce any electron transitions in atoms. In laboratory experiments an electric field leading to the complete mixing of $2s$, $2p$ states in hydrogen atom produces the Stark shift, which is much smaller than the Lamb shift. This circumstance allows the measurements of the Lamb shift with a high accuracy \cite{mohr,hilley}. The one-photon dipole emission (\ref{3}) occurs at the resonant frequency $\omega_{\overline{2s}1s} = \omega_{2s1s}+\Delta E_{2s}^{\rm Stark} \approx \omega_{2s1s}$, which is Lamb shifted from Ly$_{\alpha}$ transition frequency $\omega_{\overline{2p}1s}\approx \omega_{2p 1s}$. With the growth of strength of the electric field the intensity of emitted photons at the frequency $\omega_{2s1s}$ increases and reaches the Ly$_\alpha$ value at the field strength corresponding to the complete mixing of $2s$, $2p$ states. The difference of emission frequencies can serve as a tool for distinguishing of mixing and RS processes in laboratory experiments, see Fig.~\ref{fig1}.

To make this physical picture clearer, the RS process should be considered in the case when the excited state $a$ is mixed (we denote it by $\overline{a}$). Formally, it can be obtained by the substitutions $a\rightarrow \overline{a}$ and $E_a\rightarrow E_{\overline{a}} = E_a+\Delta E_a^{\rm Stark}\approx E_a$, $\Gamma_{\overline{a}}=\Gamma_a+\Gamma_a^{\rm BBR-QED}$, $W_{a f}\rightarrow W_{\overline{a} f}(\omega)$ into Eq. (\ref{12}). Then in the resonant approximation we find
\begin{eqnarray}
\label{20}
d{\it w}_{\overline{a}f}(\omega)=\frac{1}{2\pi}\frac{dW_{\overline{a} f}(\omega_{\overline{a}f})}{(E_{\overline{a}}-E_f-\omega)^2+\frac{1}{4}\Gamma_{\overline{a}}^2}.
\end{eqnarray}
In this expression the resonance frequency $\omega_{\overline{a}f}\approx \omega_{af}$. Thus the presence of static electric field does not change the resonant character of RS effect.

Therefore, there are two independent processes. The first one is given by Eqs. (\ref{18}), (\ref{19}), when the photon emission occurs at the resonant frequency $\omega_{a f}$. The second one can be obtained with QED description. It can be characterized by the one-photon emission from the field-modified level $\overline{2s}$, given by Eq. (\ref{20}) with the photon emission frequency $\omega_{\overline{a}f}$. In the case of two neighbouring $2s$ and $2p$ states (which energies are equal in non-relativistic limit) the frequencies of these two emission lines differ by the Lamb shift:
\begin{eqnarray}
\label{21}
\omega_{2p 1s}-\omega_{\overline{2s}1s} = E_{2p}-E_{2s} +\delta E_{2p}^L - \delta E_{2s}^L-\Delta E^{\rm Stark}
\nonumber
\\
=\delta E_{2p}^L - \delta E_{2s}^L-\Delta E^{\rm Stark}\approx \Delta E_{2s 2p}^L,\qquad
\end{eqnarray}
where $\delta E_{a}^L$ denotes the Lamb shift of the state $a$. In total, the situation with $ 2p $ and $ 2s $ levels in the BBR field looks as follows. Both levels are broadened by RS and mixing effects. The broadening of $ 2p $ level modifies Ly$_{\alpha} $ spectral emission line but does not change the emission frequency. The broadening of $ 2s $ level leads to the arrival of the new spectral line (one-photon $ \mathrm{E1} $ transition) with the frequency Lamb-shifted from Ly$_{\alpha} $. This happens exclusively due to the mixing effect, not by RS. The broadening effects produced by the mixing are much stronger than the broadening effects produced by RS. 

The numerical calculations of $\Gamma_a^{\rm BBR-QED}$ \cite{solovyev} show that the mixing effect is dominant in comparison to the RS process, see Table~\ref{width1}.

\begin{widetext}
	\onecolumngrid
	\begin{table}[h]
		\caption{\small{Numerical values $\Gamma_a^{\rm BBR-QED}$  for $2p$ and $2s$ states in hydrogen atom in $\mathrm{s}^{-1}$ (the last two columns) for different values of radiation temperature $T$ in Kelvin (first column). The corresponding $\Gamma^{\rm BBR-RS}_{2p}$ and $\Gamma^{\rm BBR-RS}_{2s}$ in $\mathrm{s}^{-1}$ are listed in second and third columns. The number in parentheses indicates the power of ten.}}
		\label{width1}
		\begin{tabular}{  c   c   c  c   c }
			\hline
			\hline
			$T,\mathrm{K}$ & $ \Gamma^{\mathrm{BBR-RS}}_{2p} $ & $\Gamma^{\mathrm{BBR-RS}}_{2s}  $ & $ \Gamma^{\mathrm{BBR-QED}}_{2p} $ & $ \Gamma^{\mathrm{BBR-QED}}_{2s} $   \\
			\hline
			$3$     & $4.782(-8)$ & $1.434(-7)$ & $0.475$     & $1.42$  \\
			$300$   & $4.743(-6)$ & $1.422(-5)$ & $3.572(3)$  & $1.070(4)$ \\
			$1000$  & $0.033$ & $2.023(-2)$ & $5.265(4)$  & $1.208(5)$     \\
			$2000$  & $1.916(3)$ & $11.783(2)$ & $7.134(8)$  & $1.207(8)$  \\
			$3000$  & $7.583(4)$ & $470.062(2)$ & $2.759(10)$ & $4.651(9)$  \\
			$5000$  & $1.522(6)$ & $967.091(3)$ & $5.198(11)$ & $8.760(10)$ \\
			\hline
			\hline
		\end{tabular}
	\end{table}
	\twocolumngrid
\end{widetext}

In following the BBR-induced level mixing effect (\ref{5}) is discussed in application to the astrophysical investigation of the cosmological recombination epoch of the early universe (in SI units). The corresponding contribution can be evaluated similarly for the level mixing in helium atom caused by the spin-orbit interaction \cite{Dubr,somebody}. Within the 'three-level' approach \cite{seager2000} only the emission line corresponding to the one-photon decay in RS $2p\rightarrow 1s+\gamma(\mathrm{E1})$ together with the two-photon decay of the $2s$ state in hydrogen $2s\rightarrow 1s+2\gamma(\mathrm{E1})$ are taken into account. According to the discussion above the additional electric dipole decay channel $\overline{2s}\rightarrow 1s+\gamma(\mathrm{E1})$ should be included into the rate equations. 

The latter can be transformed to the differential equation for the ionization fraction $x_{e} = n_e/n_{\rm H}$, where $n_e$ is the free electron number density and $n_{\rm H}$ is the total number density of hydrogen atoms and ions. The time evolution of the density number of free electrons in a homogeneous, isotropic expanding universe can be described by the following differential equation
\begin{eqnarray}
\label{22}
\frac{d{n}_{e}}{dt}=-\sum\limits_{nl}\left(\alpha_{\mathrm{H},nl}n_{e}n_{p}-\beta_{\mathrm{H},nl}n_{nl}\right)-3n_{e}\;,
\end{eqnarray}
where $n_{nl}$ is the number density of neutral hydrogen in the state with principal quantum number $n$ and orbital momentum $l$, $n_{p}\simeq n_{e}$ is a number density of protons, $\alpha_{\mathrm{H},nl}$ is the recombination coefficient for the level $nl$ and $\beta_{\mathrm{H},nl}$ is the corresponding ionization coefficient. The last term in Eq. (\ref{22}) describes the decreasing of number density $n_e$ due to the cosmological expansion. The redshift $z$ is related to time by the expression $dz/dt=-(1+z)H(z)$, where $H(z)$ is the Hubble factor \cite{seager2000}. The radiation temperature $T_{R}$ is related to redshift as $T_{R}=T_{0}(1+z)$, where $T_{0}=2.725$ K is the present Cosmic Microwave Background (CMB) temperature.

Then Eq. (\ref{22}) can be rewritten in terms of ionization fraction $x_{e}$ with the notations $x_p = n_p/n_{\mathrm{H}}$ and $x_{2s}=n_{2s}/n_{\mathrm{H}}$:
\begin{eqnarray}
\label{23}
\frac{d{x}_{e}}{dt}=-\left(\alpha_{\mathrm{H}}x_{e}x_{p}-\beta_{\mathrm{H}}x_{2s}\right)\equiv J_{\mathrm{H}},
\end{eqnarray}
where $\alpha_{\mathrm{H}}$ and $\beta_{\mathrm{H}}$ are the total coefficients of recombination and ionization, respectively. Assuming that all the uncompensated transitions to the ground state $J_{\mathrm{H}}$ proceed via the two-photon decay $2s\rightarrow 1s+2\gamma(\mathrm{E1})$ and escape of Ly$_\alpha$ photons $2p\rightarrow 1s+\gamma(\mathrm{E1})$ due to the cosmological expansion \cite{peebles,zeld} we arrive at the balance condition
\begin{eqnarray}
\label{25}
J_{\mathrm{H}}=J_{2s}^{\mathrm{E1E1}}+J_{2p}^{\mathrm{E1}},
\end{eqnarray}
where $J_{2s}^{\mathrm{E1E1}}$ and $J_{2p}^{\mathrm{E1}}$ are the corresponding uncompensated transition rates.

Following to derivation of differential equation for the ionization fraction $x_{e}$ \cite{seager2000} we can introduce the contribution $J^{\mathrm{E1}}_{2s}$ for the one-photon transition rate $\overline{2s}\rightarrow 1s+\gamma(\mathrm{E1})$:
\begin{eqnarray}
\label{26}
\tilde{J}_{\mathrm{H}}=J^{\mathrm{E1E1}}_{2s} +J^{\mathrm{E1}}_{2p}+J^{\mathrm{E1}}_{2s}.
\end{eqnarray} 
The contribution of $J^{\mathrm{E1}}_{2s}$ can be written in the same form as $J^{\mathrm{E1}}_{2p}$ \cite{seager2000}:
\begin{eqnarray}
\label{27}
J_{2s}^{\mathrm{E1}}=P_{2s1s}A_{\overline{2s}1s}\left(x_{2s}-\mathrm{exp}\left( -\frac{E_{2s}-E_{1s}}{k_{B}T}\right)x_{1s}\right).\qquad
\end{eqnarray}
The two terms in right-hand side of Eq. (\ref{27}) represent the difference between forward and backward one-photon transitions $\overline{2s}\leftrightarrow 1s$. The Einstein coefficient $A_{\overline{2s}1s}$ is defined as the partial transition rate in Eq. (\ref{5}), i.e. $A_{\overline{2s}1s}=\Gamma_{a a_0}^{\rm BBR-QED}$, where only one term from the sum over $n$ is retained with $n=a_0=2p$ and $a=2s$.

The Sobolev escape probability $P_{2s1s}$ and optical depth $\tau_{2s}$ can be written as \cite{seager2000}
\begin{eqnarray}
\label{28}
P_{2s1s}=\frac{1-e^{-\tau_{2s}}}{\tau_{2s}},
\\
\label{29}
\tau_{2s}=\frac{A_{\overline{2s}1s}n_{1s}c^{3}}{8\pi H(z) \nu_{2s1s}}\frac{g_{2s}}{g_{1s}}.
\end{eqnarray}
Here $g_{2s}$ and $g_{1s}$ are the statistical weights of the states $2s$ and $1s$, respectively, $ \nu_{2s1s} $ is the corresponding transition frequency. Insertion of (\ref{26})-(\ref{29}) into Eq. (\ref{23}) gives the differential equation for the variable $x_{e}$. 

Then the modified equation for the ionization fraction $x_{e}$ with respect to the redshift $ z $ is
\begin{eqnarray}
\label{30}
\frac{d\tilde{x}_{e}}{dz}&=&C_{\mathrm{H}}\frac{\left(\alpha_{\mathrm{H}}n_{e}\tilde{x}_{e}-\beta_{\mathrm{H}}\mathrm{exp}\left(-\frac{\Delta E_{21}}{k_{B}T}\right)\left(1-\tilde{x}_{e}\right)\right)}{H(z)(1+z)},\,\,\,
\\
C_{\mathrm{H}}&=&\frac{\frac{g_{2p}}{g_{1s}}A^{\mathrm{r}}_{2p1s}+\frac{g_{2s}}{g_{1s}}A^{\mathrm{r}}_{\overline{2s}1s}+A_{2s1s}}{\beta_{\mathrm{H}}+\frac{g_{2p}}{g_{1s}}A^{\mathrm{r}}_{2p1s}+\frac{g_{2s}}{g_{1s}}A^{\mathrm{r}}_{\overline{2s}1s}+A_{2s1s}}.
\end{eqnarray}
Here we have used the short notation for the effective coefficient $A^{\mathrm{r}}_{2s(2p)1s}\equiv P_{2s(2p)1s}A_{\overline{2s}(2p)1s}$. Thus, an additional decay channel of the $\overline{2s}$ level has arised in $C_{\mathrm{H}}$, that gives the difference with the standard case Eq. (\ref{25}). 

\begin{figure}
\caption{\small{Ionization fraction $x_{e}$ as a function of redshift $z$. The dotted-green line corresponds to the LTE case (Saha equation), the dashed-yellow line is given by the evaluation of 'ordinary' rate equation and the solid-blue line represents ionization fraction with the account for BBR-induced level mixing.}}
\centering
\includegraphics[scale=0.55]{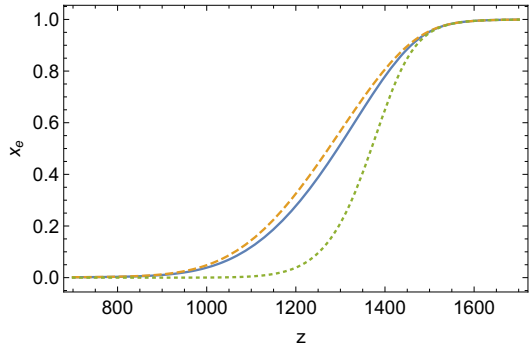}
\label{fraction}
\end{figure}

The ionization fraction $x_{e}$ was evaluated with the use of {\it Mathematica} code. All necessary cosmological parameters are taken from \cite{Planck}. The corresponding graph is presented in Fig.~\ref{fraction}, where the function $x_{e}(z)$ for the case of local thermodynamic equilibrium (LTE), i.e. evaluated with the use of Saha equation, is depicted as a dotted-green line. Evaluation of 'ordinary' rate equation \cite{seager2000} is shown by the dashed-yellow line and solid-blue line represents ionization fraction with the account for BBR-induced level mixing. 

\begin{figure}[hbtp]
\caption{Relative difference $\Delta x_{e}/x_{e}$ as a function of redshift $z$. }
\centering
\includegraphics[scale=0.55]{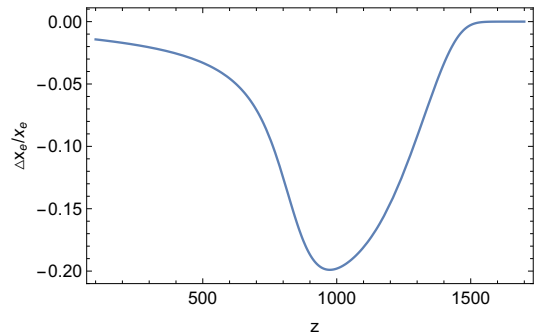}
\label{recf}
\end{figure}

We find a significant influence of the BBR-induced mixing effect on the ionization fraction in the cosmological recombination epoch of the early universe. However, the period of recombination is almost the same. Thus, the possible modification of the CMB temperature fluctuations map can be expected in the far tail of multipole expansion. The relative difference between ionization fraction from Eq. (\ref{30}) and calculated within the 'ordinary' approach $\Delta x_{e}/x_{e}\equiv (\tilde{x}_{e}-x_{e})/x_{e}$ is presented in Fig. (\ref{recf}). It is shown that the mixing effect is important during the period of cosmological recombination and reaches $20\%$ at $z\approx 1000$. Therefore the contribution of level mixing effect should be taken into account in detailed investigation of cosmological recombination epoch. It is important that the existence of the level mixing effect can be tested in laboratory experiments as it was discussed above. Hence the laboratory studies may give an access to the details of cosmological recombination.

\section*{Acknowledgements}
T. Z. acknowledges  German-Russian Interdisciplinary Science Center (G-RISC). D.S. acknowledges support from TU Dresden (DAAD Programm Ostpartnerschaften).

\bibliographystyle{apsrev4-1}

\end{document}